\documentclass[aps,twocolumn,prb,superscriptaddress]{revtex4}
\usepackage{verbatim}
\usepackage{amsfonts}
\usepackage[final,hiresbb]{graphicx}
\usepackage{amsmath}
\usepackage{amssymb}
\usepackage{amstext}

\usepackage{color}
\definecolor{Green}{RGB}{0,204,102}
\definecolor{Purple}{RGB}{102,0,255}
\definecolor{Blue}{RGB}{51,153,255}
\definecolor{Red}{RGB}{201,010,010}

\begin{document}

\title{Confinement of Vibrational Modes in Nanocrystalline Silicon}

\author{Luigi Bagolini}
\affiliation{International School for Advanced Studies, Trieste, Italy}
\author{Alessandro Mattoni}
\affiliation{Istituto Officina dei Materiali del CNR (CNR-IOM) SLACS Cagliari, Cittadella Universitaria, I-09042, Monserrato (Ca), Italy}
\author{Mark T. Lusk}
\affiliation{Department of Physics, Colorado School of Mines, Golden, CO 80401, USA}


\begin{abstract}
It is possible to confine vibrational modes to silicon nanocrystals by encapsulating them within hydrogenated amorphous silicon. This is not because of the small impedance mismatch between materials but, rather, is due to higher order moments in the distribution of density and stiffness in the amorphous phase--i.e. it is a result of material substructure. The concept is elucidated using an idealized one-dimensional setting and then demonstrated for a realistic nanocrystalline geometry. Beyond the immediate focus on amorphous encapsulation, this offers the prospect of specifically engineering higher order property distributions as an alternate means of managing phonons. The approach could be applied, for instance, to design deterministically ordered materials which exploit this means of control. 

\end{abstract} 

\maketitle
\section{Introduction}
Phonon modes can become spatially confined when the allowed vibrational frequencies in one region are not supported by the surrounding medium. A focus on such{ \em phononic gap states} or {\em vibrational stop bands}\cite{TamuraS.andWolfe1987,Santos1992,Koblinger1987a} is often motivated by a desire to manage the flow of heat or to control electronic dynamics that are coupled to phonons. The underlying physics is completely analogous to the electromagnetic effect associated with periodic structures composed of materials with differing dielectric properties. In that case, stop bands are formed in the electronic band structure preventing the propagation of electromagnetic waves within certain frequency ranges\cite{Jensen2003,Sigmund2003} which, of course, has led to the substantial field of photonic band gap materials\cite{Yablonovitch1987,John1987}.

Perhaps the simplest setting in which phonon confinement is exhibited is the one-dimensional superlattice composed of alternating layers of crystalline materials with differing elastic stiffnesses~\cite{Richter1987}. Splitting of folded bands at the center and edges of the Brillouin Zone (BZ)\cite{Yu2010} produces frequency ranges for which the composite structure does not support vibrations--i.e. for which there are no vibrational modes at all. Just as in photonic systems, this behavior can be interpreted in terms of Bragg reflections~\cite{TamuraS.andWolfe1987}. Confinement to the stiffer material does occur, though, for modes with a vibrational frequency greater than the maximum of the softer material by itself~\cite{Jensen2003}. A difference in mass density can elicit the same effect. Modes beyond this critical frequency exhibit an evanescent character in the softer material in which the mode envelope decays exponentially with distance from the interfaces~\cite{Deymier2013}. 

Such critical limit behavior has been computationally explored in superlattices composed of alternating layers of crystalline and amorphous silicon~\cite{Feldman2004a, Wu1996}.  In these nanocrystalline materials, the amorphous layer is less dense and softer than its crystalline counterpart, important because it is the average properties of each phase that determine in which confinement will occur beyond a critical frequency. It was also noted in passing, though, that there are vibrational frequencies below this critical point for which modes are confined to the crystal. This is a striking departure from bi-crystalline materials where no such confinement occurs. Motivated by this observation, we have sought a way of explaining how it is possible that amorphous silicon can cause phonons to be trapped within an encapsulated silicon nanocrystal. As will be elucidated here, such behavior can be elicited in superlattices in which one phase is itself composed of two or more crystals--i.e. that at least one of the materials has a substructure. From this perspective, the amorphous phase of nanocrystalline silicon may be viewed as composed of many, nearly degenerate crystalline phases that collectively satisfy this requirement.
Beyond the basic science, this could have important technological applications for nanocrystalline multi-functional materials\cite{Pizzini2006, Fields2014} and ultra thin silicon membranes\cite{membranes} in which phonons are managed to better assisted charge transport, engineering thermal condition and thermoelectric behavior\cite{thermoe} and modify localized phonon populations in order to slow the cooling of excited electrons and holes. 

\begin{figure}[hptb]\begin{center}
\includegraphics[width=0.48\textwidth]{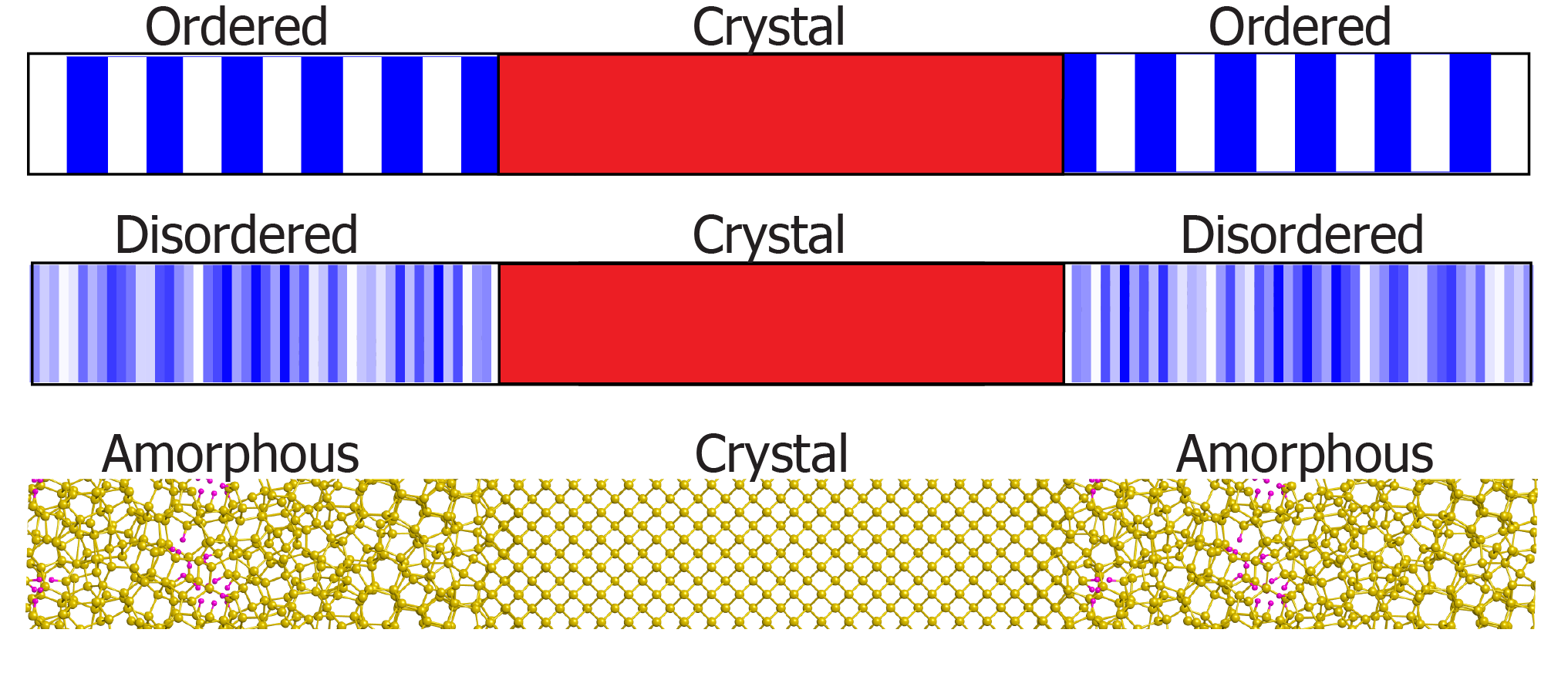}
\caption{
{\it Crystals with Structured Encapsulating Matrices.} (a) Disordered matrix, (b) ordered matrix, (c) implementation in nanocrystalline silicon.}
\label{geometry}
\end{center}\end{figure}

Unlike earlier studies, our approach starts with an explicit recognition that it is not physically relevant to carry out band structure analysis above the acoustic range for materials with a composition that includes an amorphous region. We instead focus on crystalline material encapsulated within a finite matrix. This surrounding material need not be amorphous to induce intermediate zones of crystalline confinement; it is sufficient that it has an internal substructure. This may involve a distribution of material properties, as is found in amorphous materials, or the phase may itself be a superlattice but at a finer length scale. Both are shown Fig. \ref{geometry}, as idealized one-dimensional structures, along with a nanocrystalline silicon slab that elicits the same sort of behavior. The encapsulating phases will be referred to as either {\em ordered} or {\em disordered} matrices. 

The idealized structures of Figs. \ref{geometry}(a, b) are used to elucidate the basic physics. In particular, we show that it is possible to have phonon modes confined to the crystalline phase within isolated frequency ranges that occur throughout the vibrational spectrum. These gaps, and the degree of crystalline confinement, can be tailored by engineering the distribution, as opposed to the mean, of the substructure material properties. This is possible because the substructured materials, whether an internal superlattice or amorphous in nature, have gaps in their vibrational spectra. 

Attention is then turned to the more realistic computational setting of three-dimensional, nanocrystalline silicon, in which a thin crystalline slab is embedded within a hydrogenated amorphous matrix shown in Fig. \ref{geometry}(c). The general principles and insights gleaned from the idealized 1-D chain of particles are found to have direct analogs in this technologically important form of condensed matter. In particular: (1) amorphous silicon exhibits gaps in its vibrational spectrum attributable to a distribution of stiffness and mass density; and (2) there exist vibrational modes confined to the crystalline silicon which correspond to the gaps in stand-alone amorphous silicon.

\section{Methodology}

The idealized, one-dimensional substructured superlattices of Fig. \ref{geometry}(a, b) are modeled as a harmonic chain of sites separated by a uniform distance, $a$. Each site, $n$, has mass, $\mu_n$, connected to the neighbor at its right with a spring of stiffness, $\alpha_n$. The displacement of site $n$ is given by $u_{n}$. With the left and right masses fixed, the equations of motion for the remaining sites are therefore: 
\begin{equation}
\mu_n \ddot{u}_{n} = \alpha_n u_{n\text{+}1} - (\alpha_n+\alpha_{n\text{-}1})u_{n} + \alpha_{n\text{-}1} u_{n\text{-}1} .
\end{equation}
Phonon modes, $j$, of frequency $\omega_j$, are sought so that ${u}^{j}_{n}(t) = c_n^j \mathrm{e}^{-\omega_j t}$ and
\begin{equation}
\mu_n \omega_j^2 c_n^j = -\alpha_n c^{j}_{n\text{+}1} + (\alpha_n+\alpha_{n\text{-}1})c^{j}_{n} - \alpha_{n\text{-}1} c^{j}_{n\text{-}1} .
\end{equation}
Introduction of mass-weighted coordinates, $d_n^j:= c_n^j \: \mu_n^{1/2}$, casts the equations into an eigenvalue problem for the Hermitian dynamical matrix, $[\Lambda]$:
\begin{eqnarray}
[\Lambda]\underbar{d}^j = \omega_j^2 \underbar{d}^j .
\end{eqnarray}
A numerical eigenvalue routine was then used to obtain the mass-weighted modes, $\underbar{d}^j$, and associated vibrational frequencies, $\omega_j$. For the system with the ordered matrix substructure of Fig. \ref{geometry}(a), there are 101 crystal sites (red) with 40 substructure sites on either side (alternating blue and white layers). The stiffnesses in the substructure regions alternate in 10-site increments between $0.1 \alpha$ and $1.9 \alpha$ with $\alpha$ the stiffness of the crystal. For the system with the disordered matrix substructure of Fig. \ref{geometry}(b), there are 201 crystal sites with 100 substructure sites on either side. The stiffnesses were constructed, using a uniform random number generator, to have values between $0 \alpha$ and $10 \alpha$. For both geometries, the encapsulating substructures are mirrored so that there is an axis of stiffness symmetry. This makes the analysis results particularly simple without loss of physicality. In both cases, all sites have the same mass, $\mu$, and a characteristic frequency of $\omega_0 = \sqrt{\alpha/\mu}$ is used to normalize the frequency spectra. 

The degree of confinement of mode $j$ to either phase is introduced through projections onto the matrix (M) and crystalline (C) regions:
\begin{equation}
\chi_M^j = | \underbar{M}\cdot \underbar{d}^j |^2 \quad ,  \quad \chi_C^j = | \underbar{C}\cdot \underbar{d}^j |^2 
\label{confine}
\end{equation}
where $M_n = 1$ if site $n$ is in the matrix and is zero otherwise. The reverse defines the occupancy vector, $C$. 

Molecular dynamics (MD) calculations were carried out on mixed crystalline-amorphous (c-a) systems with the phase interfaceperpendicular to the $\hat{z}$ crystallographic direction as shown in Fig. \ref{geometry}(c). A 320-atom cell, composed of a rectangular grid of 2x2x28 conventional cells, was encapsulated within a 300-atom a-Si matrix of the same density using a procedure~\cite{Nakhmanson_2001} based on the bond-switching method of Wooten, Winer, and Weaire~\cite{WWW_1985}.  The procedure begins with a nanocrystal (NC) surrounded by a periodic region in which  Si atoms have been placed randomly to give a prescribed mass density. A subsequent sequence of Monte Carlo anneal and quench steps causes the assembly to evolve into a well-relaxed composite aSi/NC structure with a standard deviation in bond lengths and angles as small as 0.073 \AA \,  and $7^{\circ}$, respectively.  This results in relatively localized electronic levels (defects) in the amorphous structures~\cite{Nakhmanson_2001}. Density Functional Theory (DFT), within a generalized gradient approximation, was then employed to generate an associated electronic structure. This allowed localized defect states to be identified and removed using an accepted hydrogenation procedure (8\% H)~\cite{Allan_1998}. This methodology and its application to electronic structure calculations are detailed elsewhere\cite{Bagolini:2010jf,Lusk2014,Bagolini2014}.

The nano crystalline system was cut along a (100) plane at the center  of the amorphous layer with a vacuum larger than cutoff distance for Tersoff interactions inserted. Atomic forces were then fully minimized to obtain a crystalline Si crystal slab embedded within hydrogenated amorphous layers. It was determined that the original amorphous/crystalline interfaces were essentially unaffected during the relaxation. The result was brought into a molecular dynamics simulator where its vibrational spectrum was calculated via diagonalization of the dynamical matrix under a harmonic approximation\cite{Mattoni2016}. The interatomic forces  were obtained from the Hansen and Vogl\cite{Hansen1998} interatomic potential, derived from the Tersoff potential\cite{Tersoff1989} for silicon, which has been successfully applied to the study the microstructural evolution of hydrogenated amorphous-crystalline  silicon systems\cite{Fugallo2014}. Each vibrational mode was expressed in terms of mass-weighted atomic displacements, and the degree of confinement to either phase followed the same approach as laid out for the finite chains.

\section{Results and Discussion}

\begin{figure}[hptb]\begin{center}
\includegraphics[width=0.4\textwidth]{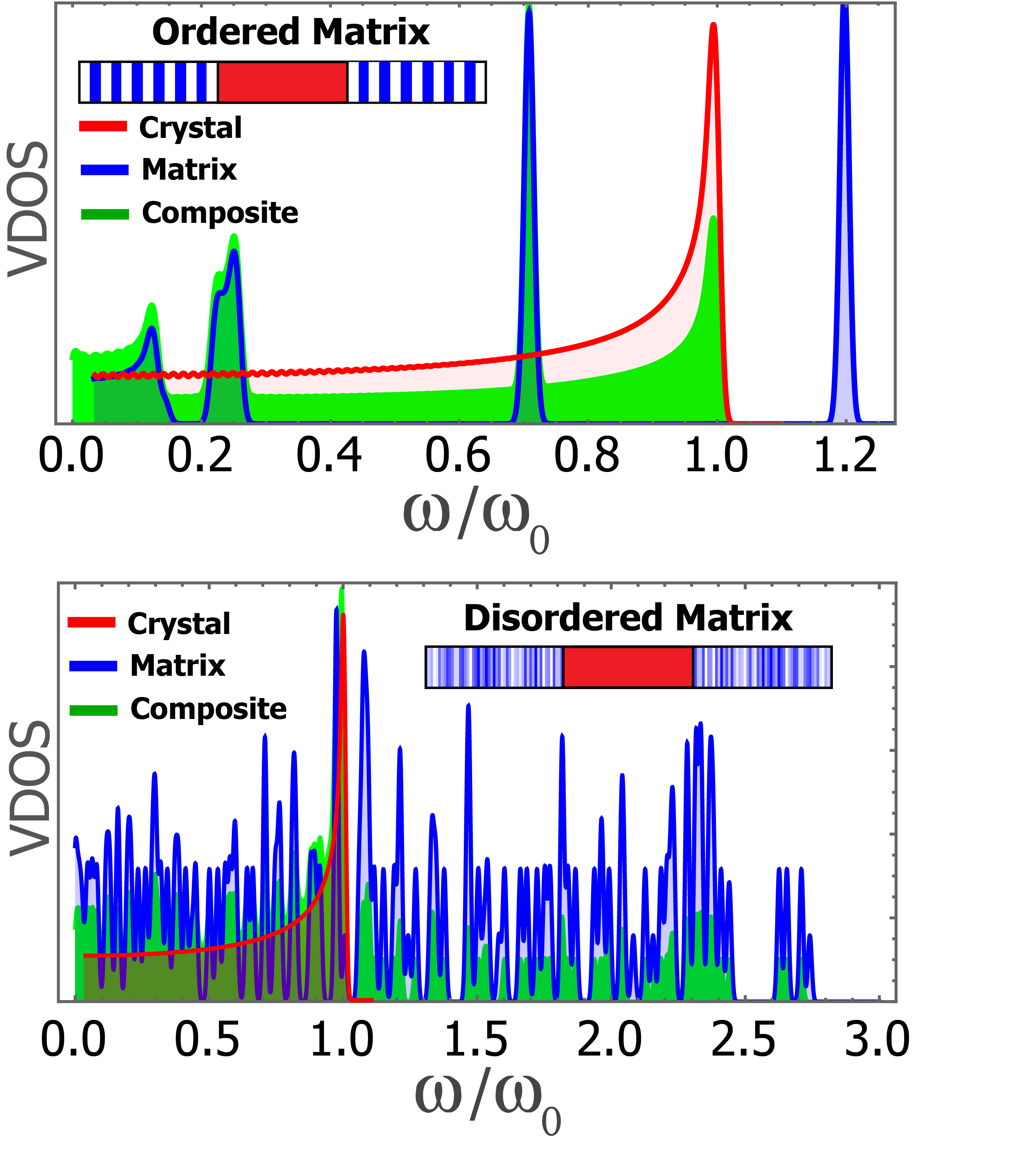}
\caption{
{\it Full Spectrum of VDOS for 1-D Geometries.} For both ordered (top panel) and disordered (bottom panel) matrix encapsulations, vibrational density of states spectra are shown for stand-alone crystal (red), stand-alone matrix (blue) and composite systems (green). Each spectrum is normalized so that its peak value is unity.}
\label{1D_spectra}
\end{center}\end{figure}

The idealized, one-dimensional chain setting was employed first to establish relationships that were subsequently tested using molecular dynamics. Several observations can be made concerning the complete vibrational density of states (VDOS) for the geometries of Fig. \ref{geometry}(a, b) which are shown in Fig. \ref{1D_spectra}. The stand-alone crystal (red, both panels) is homogeneous and so has a relatively uniform VDOS. The single peak at $\omega=\omega_0$ is analogous to the high density of states observed in a simple periodic lattice at the edge of its Brillouin zone, and there is only one branch because the primitive cell is composed of a single particle. The finite domain size only serves to spread this peak just as happens with the Fourier spectrum of a signal truncated after a finite time. 

The stand-alone ordered matrix (blue, top panel) amounts to a finite version of a superlattice, and 20 particles per primitive cell result in as many band edges and resulting broadened peaks in the VDOS. The gaps between band edges of an infinite superlattice are manifested in the finite domain as gaps in the blue VDOS.

Crucially, the stand-alone disordered matrix (blue, bottom panel) also exhibits gaps in its VDOS. This is possible any time a material is composed of more than one phase, and the disordered matrix may be viewed as a collection of many distinct materials. The random distribution of these gaps is generated by the uniform random distribution of mass values chosen for the matrix material.  As a result, there are thin frequency ranges that are not supported by the matrix but are supported by the crystal---the basis for phonon confinement. 

Because the mean value of mass and stiffness is the same in both the crystal and the random matrix,  phonon confinement in such a nanocrystalline setting can be attributed to higher order moments in material properties. In both panels, the composite VDOS reflects the character of both crystal and matrix. This is not simply a direct sum, as would be the case if the two phases were uncoupled, but regions where the matrix has a stop gap have a counterpart in the composite VDOS.

%

\begin{figure}[hptb]\begin{center}
\includegraphics[width=0.48\textwidth]{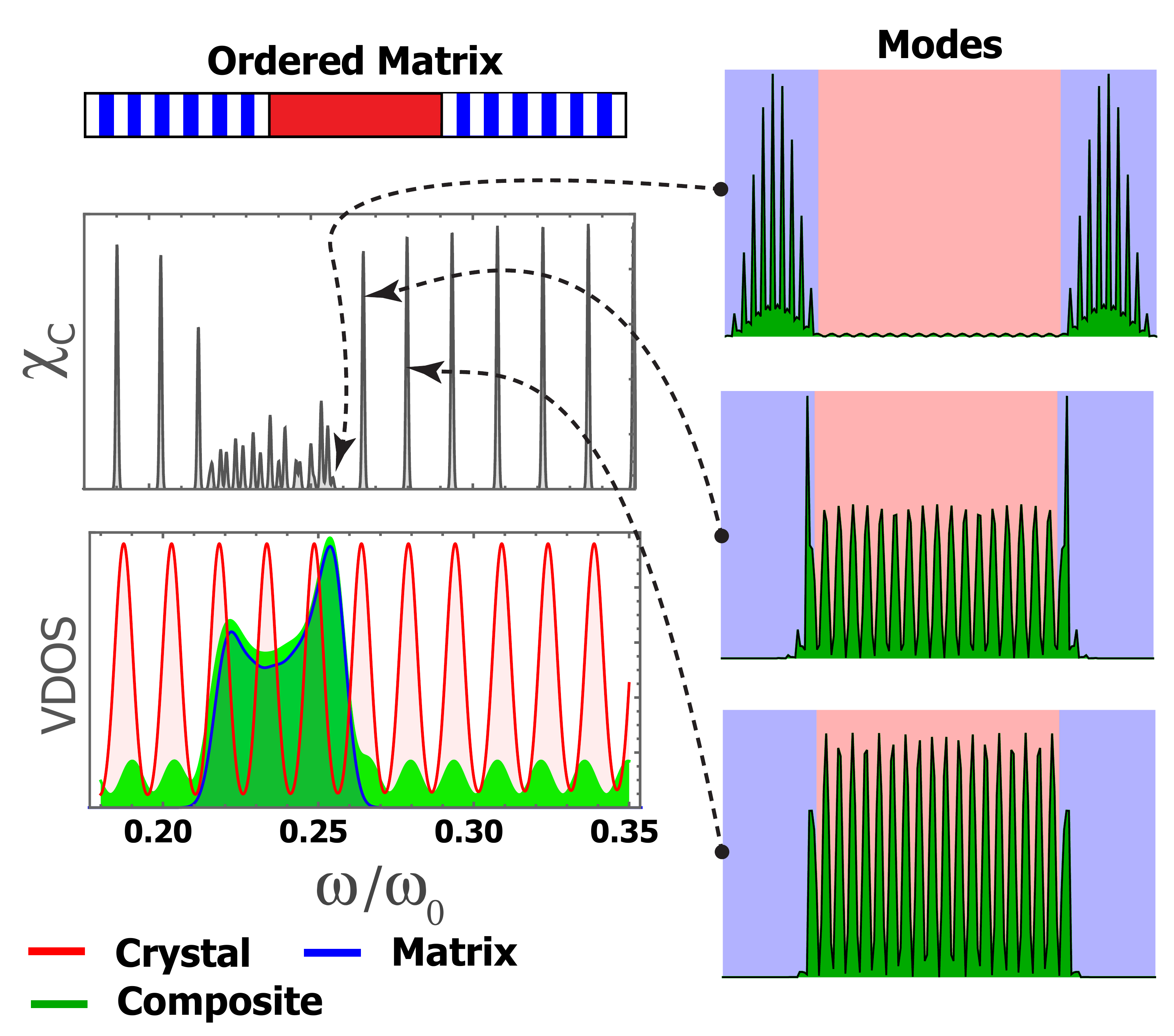}
\caption{
{\it Phonon Confinement for 1-D System with Ordered Matrix.} Vibrational density of states spectra are shown for stand-alone crystal (red), stand-alone matrix (blue) and composite systems (green). Each spectrum is normalized so that its peak value is unity. The vibrational mode plots are for the square of the amplitudes. The composite system is used to quantify the degree of confinement to the crystalline region, $\chi_C$, as defined in Eq. \ref{confine}.}
\label{1D_substructured_ordered}
\end{center}\end{figure}

Having overviewed the overall spectral character of each geometry, we now focus on very small frequency ranges in order to precisely quantify phonon confinement in the crystal. Fig. \ref{1D_substructured_ordered} shows one such zone for the ordered system, where a lower value of spectral broadening makes individual contributions apparent. Frequencies that are supported by the matrix exhibit low confinement measures, $\chi_C$,  because the vibration can spread throughout the composite, while frequencies in the stop gaps show strong crystal confinement. To further emphasize the effect, representative phonon modes are plotted at right that show confinement to the matrix (top) and crystal (middle and bottom) along with the characteristic evanescent decay in disallowed regions. 

%

\begin{figure}[hptb]\begin{center}
\includegraphics[width=0.48\textwidth]{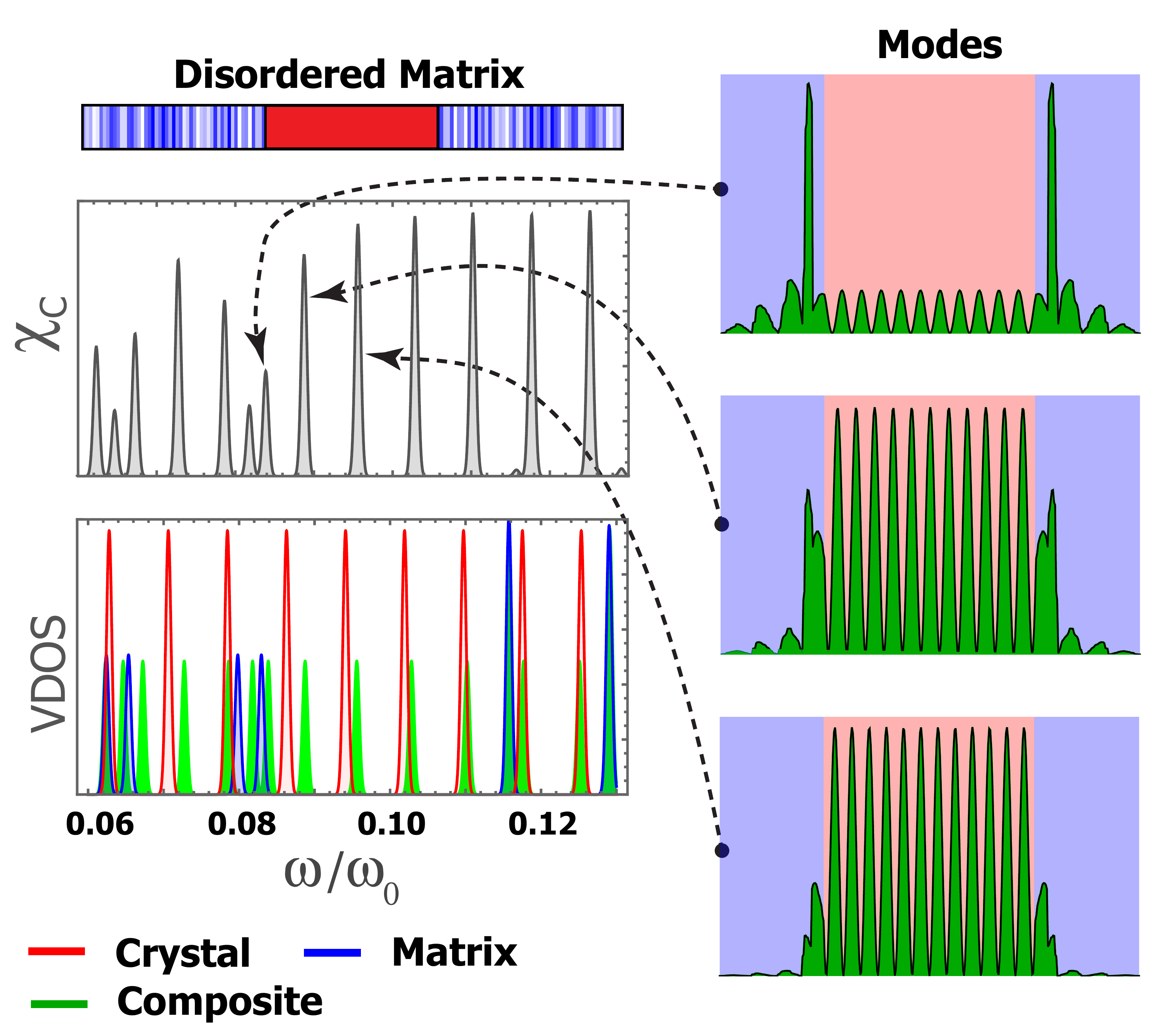}
\caption{
{\it Phonon Confinement for 1-D System with Disordered Matrix.} Vibrational density of states spectra are shown for stand-alone crystal (red), stand-alone matrix (blue) and composite systems (green). Each spectrum is normalized so that its peak value is unity. The vibrational mode plots are for the square of the amplitudes.}
\label{1D_substructured_random}
\end{center}\end{figure}

Likewise, the disordered matrix induces crystalline confinement, $\chi_C$, as quantified in Fig. \ref{1D_substructured_random}. In complete analogy to the ordered matrix, frequency ranges exhibiting confinement can be predicted from the stand-alone spectra and associated vibrational modes are strongly localized in the crystal. Since the mean stiffness and mass of the matrix is identical to that of the crystal, it is their higher order moments that elicit this behavior.

%

%
We next consider realistic crystalline-amorphous microstructures with the use of molecular dynamics.  The insights and principles obtained from the linear chain analysis are found to have direct application to hydrogenated nanocrystalline silicon of Fig.\ref{geometry}(c).  The average density and elastic stiffness of the amorphous regions are reduced by  5\% and 9\%, respectively, relative to crystalline silicon.  This is consistent with a reduction in density by 10\% in a-Si that has been experimentally measured in amorphous silicon at high hydrogenation \cite{Fahrner2013} and a  reduction of the bi-axial moduli by a 20\% reported for amorphous films\cite{Witvrouw1993}. The average density and elastic stiffness of the amorphous regions are found to be reduced by factors  0.95 and 0.91, respectively, relative to crystalline silicon.  A 
reduction of density by 10\% in a-Si has been experimentally measured in amorphous silicon at high hydrogenation \cite{Fahrner2013} and a  reduction of the bi-axial moduli by a factor 
20\% was reported in amorphous films\cite{Witvrouw1993}.

Despite the re-assuring consistency with experimental measurements, these mean-value changes to density and stiffness in amorphous silicon serve only to generate a small impedance mismatch that is not relevant to true phonon confinement. Rather, it is the substructure variations that result in such behavior, and fluctuations of stiffness and density at the atomic scale are found to have relative standard deviations of $48\%$ and $5.4\%$, respectively. The property variations, in turn, derive from variations in the bond length and dihedral angle that influence both local density and bond order.  The 1-D analysis established a link between substructure variations and phonon confinement, and analogous results were found in the nanocrystalline setting as will now be detailed.

\begin{figure}[hptb]\begin{center}
\includegraphics[width=0.48\textwidth]{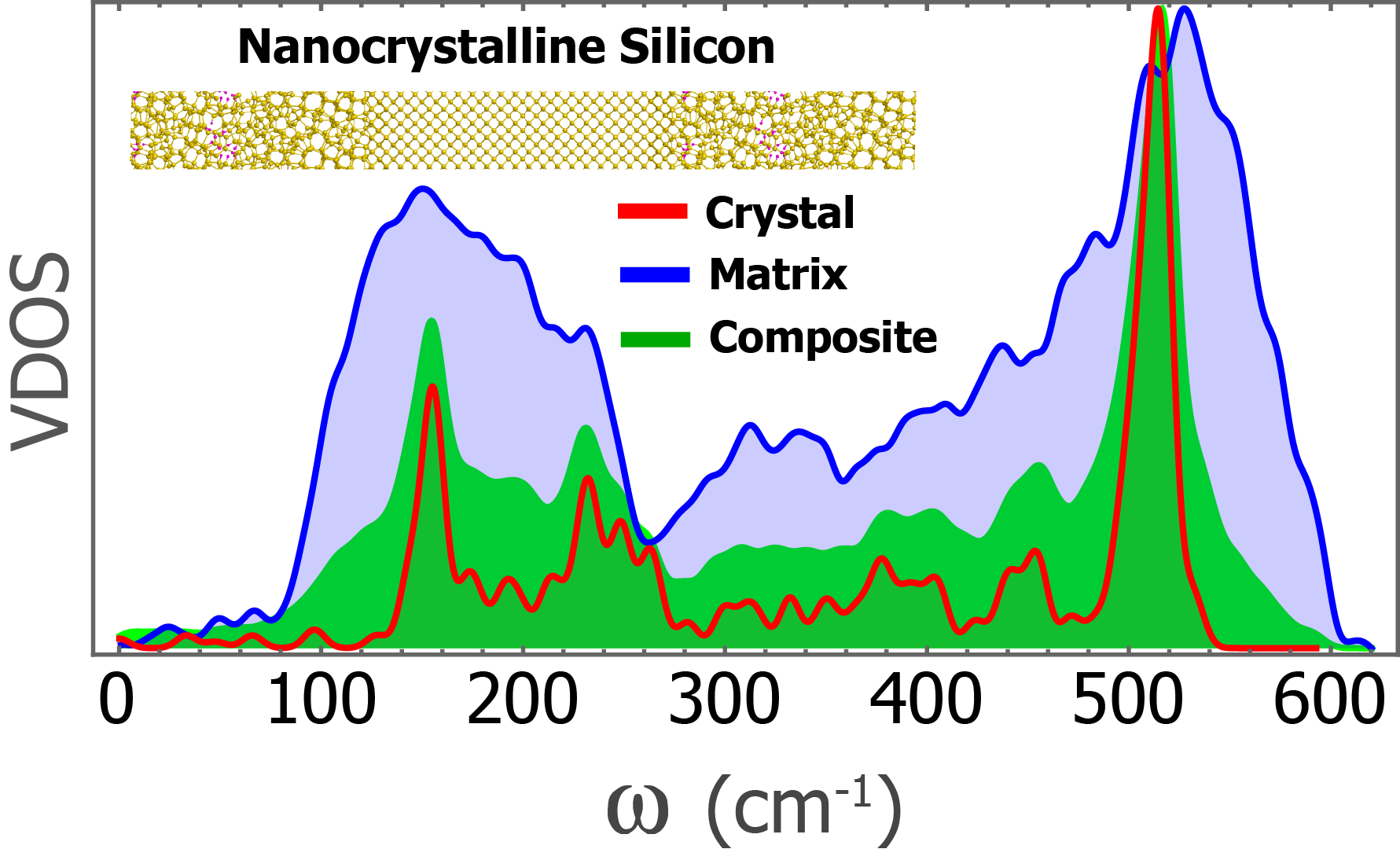}
\caption{
{\it Full Spectrum of VDOS for 3-D Nanocrystalline Silicon.} Vibrational density of states spectra are shown for stand-alone crystal (red), stand-alone matrix (blue) and composite systems (green). Each spectrum is normalized so that its peak value is unity.}
\label{MD_summary}
\end{center}\end{figure}

The non-periodic structure consists of a crystal  layer of $\sim5.5$ nm embedded into two amorphous layers of thickness $\sim4.5$ nm. As shown in Fig. \ref{MD_summary} and qualitatively consistent with a wealth of experimental 
data~\cite{Li1989,Wienkes2013,Duan2012}, the VDOS of nanocrystalline silicon consists of a broad band extending above  500 cm$^{-1}$ followed by  isolated modes  due to Si-H vibrations at higher frequencies. The figure also shows the VDOS of the stand-alone amorphous and crystalline phases which makes it clear that the behavior of the composite system can be largely anticipated from the character of its constituents. Coupling of the regions causes deviations in peak positions on the order of 20 cm$^{-1}$ and are smaller in the low frequency region (below 300 cm-1).  This is consistent with the 1-D analysis.

As shown in Fig. \ref{MD_summary}, the stand-alone crystalline silicon slab (red) exhibits broadened Van Hove peaks that correspond to the L, X, W and K symmetry points in the crystalline silicon Brillouin zone. Unlike the random distribution of properties employed in the 1-D chain of particles, there is a smooth distribution of nearly degenerate properties in the more physical 3-D atomistic setting. These stem from variations in dihedral angles and bond length that, in turn, change the local stiffness and density. The VDOS reflects this by exhibiting broadened features about crystalline accumulation points. For instance, amorphous silicon (blue) has an extended populations between 100 and 260 cm$^{-1}$ with a strong feature near 155 cm$^{-1}$ that is associated with phonons propagating along the $\langle 111\rangle$ crystallographic direction (L-point) in crystalline silicon.  The result is a broad shoulder in the range 100-200 cm$^{-1}$ with, significantly, a larger average distance between vibrational frequencies than for the crystal. These small gaps in the amorphous VDOS will, statistically speaking, often be occupied by frequencies that are supported in the crystal, and this gives the confinement sought. We choose to focus on this region in particular to examine the possibility of such confinement.

\begin{figure}[hptb]\begin{center}
\includegraphics[width=0.48\textwidth]{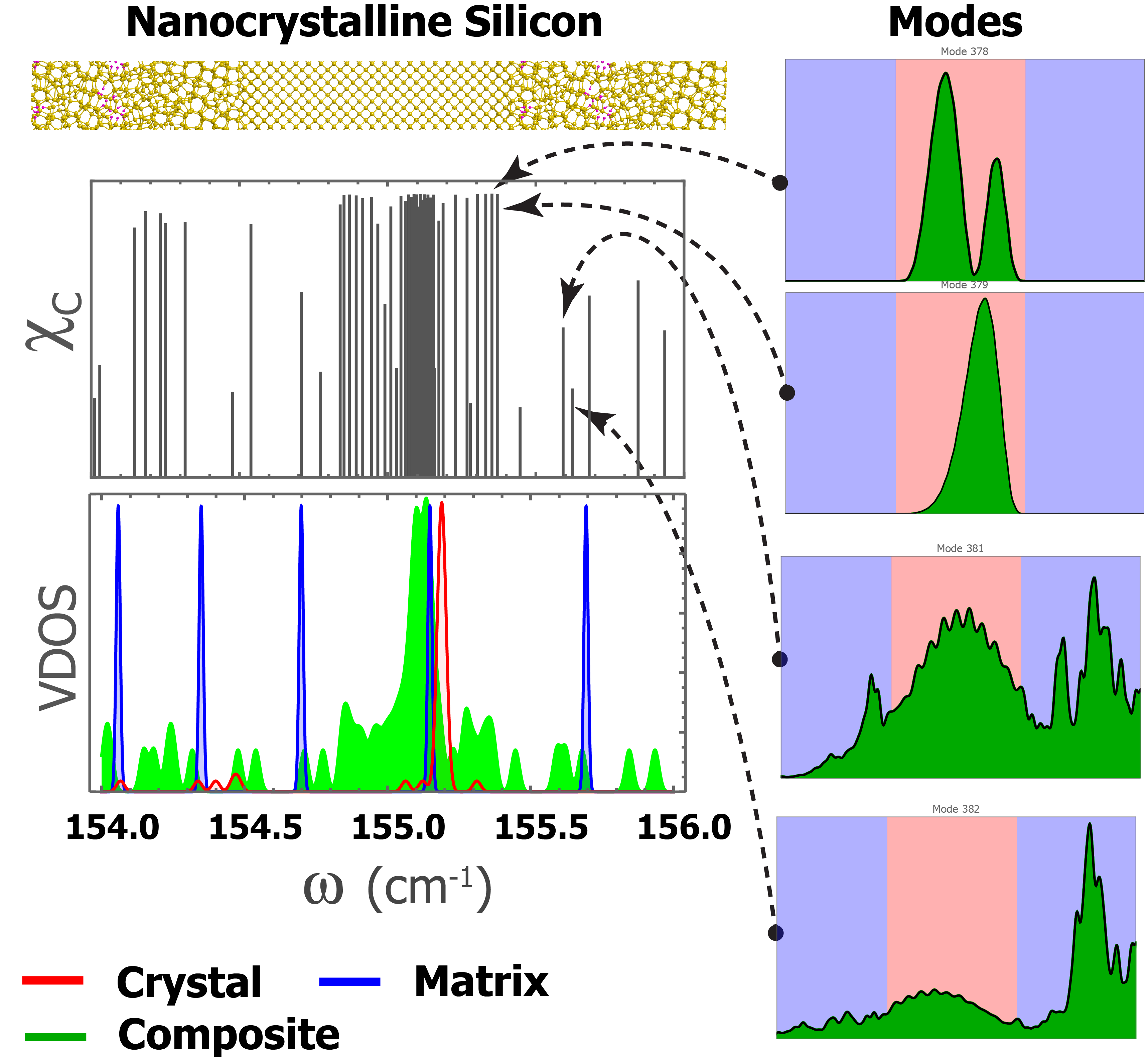}
\caption{
{\it Phonon Confinement for 3-D Nanocrystalline Silicon.} Vibrational density of states spectra are shown for stand-alone crystal (red), stand-alone matrix (blue) and composite systems (green). Each spectrum is normalized so that its peak value is unity. The vibrational mode plots are for the square of the amplitudes.}
\label{MD_detail}
\end{center}\end{figure}

In order to precisely quantify such crystalline confinement, a small region of the vibrational spectrum was chosen within the peaked zone with L-point character: 154-156 cm$^{-1}$. As shown in Fig. \ref{MD_detail}, the modes of the isolated crystalline slab form a sharp red peak at 155.19 cm$^{-1}$ that broadens in the composite (green band within 154.8--155.37 cm$^{-1}$). The amorphous levels are much more separated in energy ($\sim$ 52 cm$^{-1}$), as discussed above, and as result some crystalline levels fall within a frequency range that is not supported by the amorphous material. Plots of selected vibrational modes confirm the evanescent nature of these modes within the matrix.

While the idealized 1-D lattice helped to elucidate the origin of phonon confinement, the analysis of hydrogenated nanocrystalline silicon shows essentially the same behavior. The small distribution of dihedral angle and bond lengths in the amorphous phase cause a distribution in density and stiffness, and such a substructure can be viewed as a composition of distinct materials. When used to encapsulate a silicon crystal, the requirement for phonon confinement is satisfied and there will exists modes that are localized to the crystal.

%

\section{Conclusions}

It is possible to trap vibrational modes within silicon nanocrystals by encapsulating them within a hydrogenated amorphous silicon matrix. This is not because of the small impedance mismatch between materials but, rather, is due to higher order moments in the distribution of density and stiffness in the amorphous phase--i.e. it is a result of material substructure. This was elucidated within an idealized one-dimensional setting where it was shown that even a simple two-level variation in stiffness within the matrix is sufficient to elicit such confinement. Within the same setting, a random structure---an idealization of an amorphous matrix---was also shown to have the same behavior. A combination of Monte Carlo, Density Functional Theory, and Molecular Dynamics was then used to demonstrate that it is possible to have vibrational modes that are confined to nanocrystals of silicon encapsulated within a hydrogenated amorphous silicon matrix. This investigation focuses on the elucidation, not exploitation, of behavior attributable to substructural variations. It opens the door, though, to the prospect of specifically engineering property distributions in order to manage phonons so as to enhance or mitigate dynamics related to lattice vibration. This could be in association with amorphous phases, but deterministically ordered materials---e.g. finite zones of superlattices--may offer a more powerful setting in which to exploit this means of control. 

\section{Acknowledgments}
A.M. acknowledges financial support  by Regione Autonoma della Sardegna (CRP-24978 and CRP-18013), by Fondazione Banco di Sardegna (Project 7454, 5794), and computational support by CINECA, Italy through ISCRA Project VIPER. 

This material is based upon work supported by the U.S. Department of Energy under Award Number DE-EE0005326. This report was prepared as an account of work sponsored by an agency of the United States Government.  Neither the United States Government nor any agency thereof, nor any of their employees, makes any warranty, express or implied, or assumes any legal liability or responsibility for the accuracy, completeness, or usefulness of any information, apparatus, product, or process disclosed, or represents that its use would not infringe privately owned rights.  Reference herein to any specific commercial product, process, or service by trade name, trademark, manufacturer, or otherwise does not necessarily constitute or imply its endorsement, recommendation, or favoring by the United States Government or any agency thereof.  The views and opinions of authors expressed herein do not necessarily state or reflect those of the United States Government or any agency thereof.


\end{document}